\documentclass{article}
\usepackage{spconf,amsmath,graphicx,amssymb}
\usepackage{booktabs, threeparttable, makecell}
\usepackage{multirow,array,enumitem,adjustbox}
\usepackage[linesnumbered,ruled,vlined]{algorithm2e}
\usepackage[colorlinks=true,citecolor=green]{hyperref}
\usepackage{cleveref}

\usepackage{caption} 
\newcommand{\slfrac}[2]{\left.#1\middle/#2\right.}

\title{A comprehensive study on self-supervised distillation for speaker representation learning}
\name{Zhengyang Chen$^{1,2}$ \thanks{This work was supported in part by the China NSFC projects (Grant No. 62122050 and No. 62071288), and in part by Shanghai
Municipal Science and Technology Major Project (Grant No.
2021SHZDZX0102).}, Yao Qian$^{2}$, Bing Han$^{1}$, Yanmin Qian$^{1,*}$ \thanks{$^*$Corresponding author.}, Michael Zeng$^2$ }
\address{$^1$ MoE Key Lab of Artificial Intelligence, AI Institute
, \\
X-LANCE Lab, Department of Computer Science and Engineering, Shanghai Jiao Tong University\\
$^2$Microsoft Cognitive Services Research, USA}
\copyrightnotice{978-1-6654-7189-3/22/\$31.00~\copyright2023 IEEE}
\begin{document}
\ninept
\maketitle
\begin{abstract}
In real application scenarios, it is often challenging to obtain a large amount of labeled data for speaker representation learning due to speaker privacy concerns. Self-supervised learning with no labels has become a more and more promising way to solve it. Compared with contrastive learning, self-distilled approaches use only positive samples in the loss function and thus are more attractive. In this paper, we present a comprehensive study on self-distilled self-supervised speaker representation learning, especially on critical data augmentation. Our proposed strategy of audio perturbation augmentation has pushed the performance of the speaker representation to a new limit. The experimental results show that our model can achieve a new SoTA on Voxceleb1 speaker verification evaluation benchmark ( i.e., equal error rate (EER) 2.505\%, 2.473\%, and 4.791\% for trial Vox1-O, Vox1-E and Vox1-H , respectively), discarding any speaker labels in the training phase.

\end{abstract}
\section{Introduction}
\label{sec:intro}

Benefiting from the advantages of deep learning,  speaker verification systems have achieved great progress in the recent few years. Researchers have proposed different architectures \cite{liu2015deep,snyder2017deep,snyder2018x-vector,zeinali2019but,desplanques2020ecapa} to encode speaker information and designed different kinds of supervised loss functions \cite{wan2018generalized,deng2019arcface,xiang2019margin} to learn  discriminative speaker representations. However, deep-learning-based methods always require a lot of labeled data for training to deliver a good-performance model. However, due to the data privacy issue and labeling cost, it is challenging for us to access speaker-labeled data. On the other hand, there is a massive amount of unlabeled data on the internet.   There appeared many emerging technologies that leverage unlabeled data in the training process. 

Self-supervised learning is one of the approaches. The most commonly used method for self-supervised speaker representative learning is the contrastive-learning based method \cite{huh2020augmentation,xia2021self,zhang2021contrastive,sang2022self}, which assumes that there is only one speaker in one utterance and different utterances contain different speakers. Then the system can do contrastive learning by maximizing the similarity between positive pairs sampled from the same utterance and the discrepancy between negative pairs sampled from the different utterances. However, it cannot guarantee that different utterances contain different speakers when no speaker labels are available. Recently, researchers in the computer vision field have proposed a bootstrap-based method \cite{grill2020bootstrap}, called BYOL,  which can do self-supervised learning by only minimizing the distance between positive pairs. Following this study, another work called self-distillation with no labels (DINO) \cite{caron2021emerging} simplified the model structure of BYOL and proposed a more effective training strategy to further improve the system performance. Researchers have also applied the DINO strategy in self-supervised speaker representation learning \cite{jung2022pushing,heo2022self}.

Although the above attempts to use DINO for speaker representation learning are successful,  there are still few investigations on the characteristics of this method to further push its performance. In this paper, we have done a comprehensive study on the DINO strategy used in self-supervised speaker representation learning.  The major findings are listed as followings,
\begin{itemize}
    \item Audio augmentation by adding noise and reverberation can significantly improve the performance of the DINO-based speaker representation learning system.
    \item Among the perturbation-based augmentations, pitch perturbation can further improve the performance only when more iterations are used in the training and tempo perturbation has little effect on the system performance.
    \item When the size of the dataset is fixed, optimizing the DINO objective becomes more difficult when there are more speakers and thus the system performance degrades.
    \item The output DINO distribution can be approximately considered as a one-hot speaker label when we ignore the small values, which indicates a strong correlation with the true label.
    \item The DINO strategy shows a great superiority over the contrastive learning-based method.
\end{itemize}

Based on the above findings, we pushed the self-supervised learning performance on the Voxceleb1 evaluation benchmark to a new limit.

\section{Related Work}
\label{sec:related_work}
In recent years, researchers have made great efforts to leverage the unlabeled data in the speaker verification task. In the early work, Stafylakis et al. \cite{stafylakis2019self} trained the model to extract meaningful speaker embedding by using a feature reconstruction loss and additional phone information.
%Apart from this relatively early work, the self-supervised speaker representation learning methods can be roughly divided into two categories: the contrastive-learning-based method and the bootstrap \cite{grill2020bootstrap, caron2021emerging} based method.
%Apart from this relatively early work, 
Nowadays, the contrastive-learning-based method and the bootstrap \cite{grill2020bootstrap, caron2021emerging} based method are two dominant approaches for self-supervised speaker representation learning. 

%In the contrastive-learning-based method, the basic assumption is that there is only one speaker in one utterance and different utterances contain different speakers. With the above assumption, positive pairs can be sampled from the same utterance and negative pairs can be sampled from different utterances. 
For the self-supervised constrastive learning, Huh et al. \cite{huh2020augmentation} proposed augmentation adversarial training to help the network learn channel invariant information and Zhang et al. \cite{zhang2021contrastive} proposed a channel invariant loss to learn more robust speaker representation. Further, Xia et al. \cite{xia2021self} proposed a prototypical memory bank to enlarge the negative sample pool and avoid the latent false negative pairs. Besides, Sang et al. \cite{sang2022self} proposed a siamese network and a self-supervised regularization strategy to further improve the contrastive learning performance.

%Compared with the contrastive learning based method, only one assumption should be satisfied in the bootstrap-based method that there is only one speaker in one utterance. 
The bootstrap-based method is an emerging technology. Recent works mainly focus on DINO \cite{caron2021emerging} based bootstrap method. Heo et al. \cite{heo2022self} found that the DINO training strategy can converge better when there are fewer speakers in the training set and they proposed a curriculum learning strategy to improve the performance.
Jung et al. \cite{jung2022pushing} combined the DINO training strategy with a raw waveform-based neural network. However, compared with the contrastive-learning-based method, the bootstrap-based method is less studied in the field of self-supervised speaker representation learning. In this paper, we will give a deep investigation on applying the DINO strategy in self-supervised speaker verification task.

\section{Method}
\label{sec:dino}
In this section, we introduce the self-\textbf{di}stillation with \textbf{no} labels (DINO) based self-supervised training strategy and apply it to learn speaker representation using unlabeled data.

\subsection{Self-Distillation with No Labels}
\label{ssec:dino_system_intro}
\begin{figure}[ht!]
  \centering
  \includegraphics[width=0.48\textwidth]{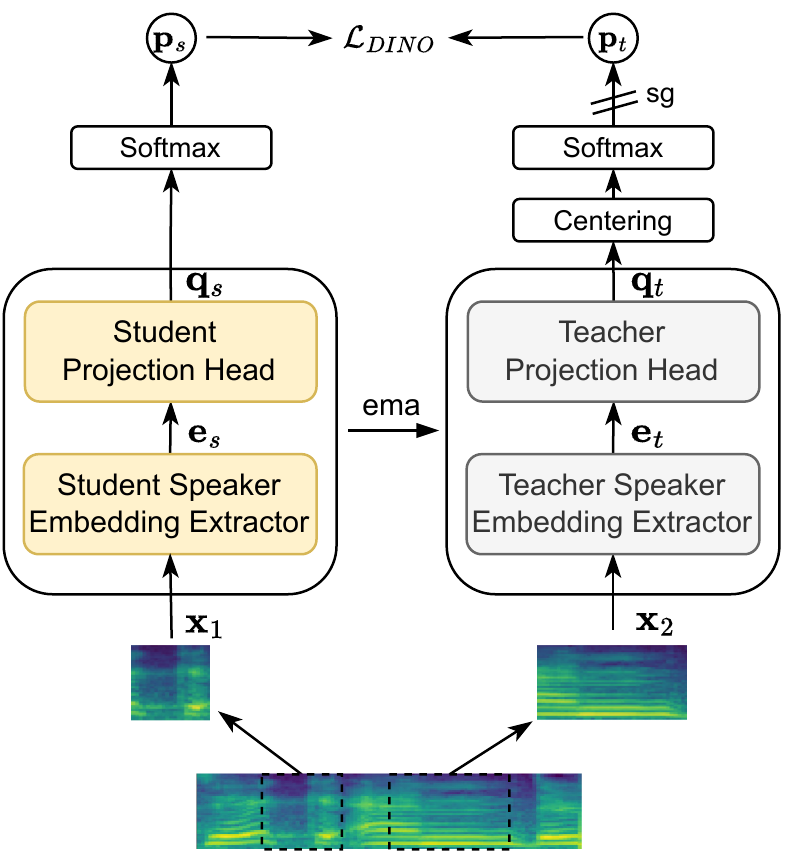}
  \caption{\textbf{Speaker representation learning with DINO.} There could be multiples segments from one utterance fed into the student or teacher network. We only plot one segment in this figure. ``sg'' denotes the stop gradient operation.}
  \label{fig:dino}
 \vspace{-10pt}
\end{figure}

As shown in Figure \ref{fig:dino}, the DINO strategy trains the model in a knowledge distillation (KD) paradigm. The output from the student network is optimized to match the output from the teacher network. 
Different from the common KD method, the teacher model in DINO is not pre-trained but derived from the student network. Besides, the student and teacher networks share the same architecture but have different parameters. Here, we denote student network as $\mathcal{F}_s$, parameterized by $\theta_{s}$ and denote teacher network as $\mathcal{F}_t$, parameterized by $\theta_{t}$. The $\theta_{t}$ is exponentially moving averaged (ema) \cite{he2020momentum} from $\theta_{s}$ in the rule $\theta_{t} \leftarrow \lambda \theta_{t}+(1-\lambda) \theta_{s}$, where $\lambda$ follows a cosine scheduler from 0.996 to 1 \cite{grill2020bootstrap} during training process.

Both student and teacher networks contain two modules, a speaker embedding extractor $g$ and a projection head $h$: $\mathcal{F} = h \circ g$. The speaker embedding extractor can be any kind of structure that can map the variable-length acoustic feature sequence, $\mathbf{x}$, to fix-dimensional speaker embedding, $\mathbf{e} = g(\mathbf{x})$. The projection head consists of some fully connected layers that maps the speaker embedding $\mathbf{e}$ to a $K$ dimensional vector, $\mathbf{q}= h(\mathbf{e})$. For the $\mathbf{q}_s$ from the student network, another softmax function will be applied to map $\mathbf{q}_s$ to a probability distribution $\mathbf{p}_s$:

\begin{equation}
\label{eq:get_prob}
\begin{split}
p_s^{(i)}=\frac{\exp \left(q_s^{(i)} / \tau_{s}\right)}{\sum_{k=1}^{K} \exp \left(q_s^{(k)} / \tau_{s}\right)}
\end{split}
\end{equation}
where $i$ is the $i$-th dimension of $\mathbf{p}_s$ and $\tau_{s}$ is the temperature to control the sharpness of the distribution $\mathbf{p}_s$. The probability distribution $\mathbf{p}_t$ from the teacher branch is calculated similarly but with a different temperature $\tau_{t}$. In addition, there is a centering operation before softmax. The centering operation is very similar to the batchnorm operation but only does the mean normalization. The centering operation normalizes the teacher output $\mathbf{q}_t$ by a mean statistic $\mathbf{c}$: $\mathbf{q}_t \leftarrow \mathbf{q}_t-\mathbf{c}$. The statistic $\mathbf{c}$ is updated during the training process with a moving average strategy:

\begin{equation}
\vspace{-10pt}
\label{eq:c_update}
\begin{split}
\mathbf{c} \leftarrow m \mathbf{c}+(1-m) \frac{1}{B} \sum_{i=1}^{B} \mathbf{q}_t
\end{split}
\end{equation}
where $B$ is the batch size in one iteration and $0<m<1$ is the momentum factor. The student network is optimized to minimize the divergence between distribution $\mathbf{p}_t$ and $\mathbf{p}_s$ measured by cross-entropy loss. Because there is no additional supervisory signal here, the teacher model can easily collapse to the trivial solution that the teacher network always outputs a uniform distribution $\mathbf{p}_t$ or there is a fixed dimension dominating $\mathbf{p}_t$. The centering operation can avoid the fixed dimension dominating problem and the sharpening operation achieved by a smaller temperature $\tau_t$ can ease the uniform distribution problem. These two tricks can cooperate to avoid the model collapse problem.

Similar to the implementation in \cite{caron2021emerging}, we sampled multiple segments (views) from each utterances. In our experiment, 
we sample $L$ long segments (global views) and $M$ short segments (local views) from each utterance. All the segments are fed into the student network and only long segments are fed into the teacher segments. The DINO loss for each utterance is calculated as:

\vspace{-5pt}
\begin{equation}
\label{eq:dino_loss}
\begin{split}
\mathcal{L}_{DINO} = \frac{1}{L(L+M-1)}
\sum_{i=1}^{L} \sum_{\substack{j=1 \\ j \neq i}}^{L+M}
\text{CrossEntropy}(\mathbf{p}_t^i, \mathbf{p}_s^j)
% &\text{CrossEntropy}(\text{softmax}(\mathcal{F}_t(\mathbf{x}_i)-\mathbf{c}), \text{softmax}(\mathcal{F}_s(\mathbf{x}_j)))
\end{split}
\end{equation}
\vspace{-5pt}

\vspace{-5pt}
\subsection{Data Augmentation for Self-Supervised Learning}
\label{ssec:data_augmentation_intro}

Many researchers \cite{huh2020augmentation,sang2022self} have shown that the data augmentation strategy is the key to the success of self-supervised speaker representation learning. The self-supervised learning methods always have the assumption that the speaker information is the only temporally consistent information in a sentence. Thus, when we want to extract some shared information from two segments of the same utterance, the shared information could be speaker information. However, the channel information is also consistent along the whole utterance. To avoid the model learning channel information, different noises or reverberations are added on the different segments from the same utterance to break the channel information consistency across different segments. Moreover, the audio perturbation \cite{yamamoto2019speaker} augmentation, which was mainly evaluated in speaker verification challenge \cite{wang2020dku,zhao2021speakin,chen2022sjtu}, is firstly tried for DINO based self-supervised speaker representation learning in our experiments.     

\vspace{-5pt}
\begin{algorithm}[!ht]
\footnotesize
 We denote the probability of additive noise ($\text{A}$) and reverberation ($\text{R}$) augmentation  as $p_\text{A+R}$, the probability of pitch perturbation ($\text{P}$) as $p_\text{P}$, and the probability of tempo perturbation ($\text{T}$) as $p_\text{T}$, respectively.\\
For the short segments, the number of segments, segment duration and segment set from each audio are indicated as $M$, $T_{short}$ and $\mathcal{X}_{short}^i$, and the counterparts of long segments are indicated as $L$, $T_{long}$ and $\mathcal{X}_{long}^i$. \\
 
 \For{ $\text{Utterance}_i \in \text{Train Set}$ }{
        Sample $q_\text{P}$ from uniform distribution $\mathcal{U}(0, 1)$\\
        
        \If{$q_\text{P} < p_\text{P}$}
            {
                Randomly select a pitch shift value $\alpha$ from $\{-200\  \text{cent}, 200\  \text{cent}\}$ \\
                $\text{Utterance}_i = \text{sox\_pitch}(\text{Utterance}_i, \alpha)$ \\
            }
        $\mathcal{X}_{short}^i = \text{random\_segments}(\text{Utterance}_i, M, T_{short}) $ \\
        $\mathcal{X}_{long}^i = \text{random\_segments}(\text{Utterance}_i, L, T_{long}) $ \\
        \For{ $\text{segment} \in \mathcal{X}_{short} \cup \mathcal{X}_{long}$ }
        {
            Sample $q_\text{T}$ from uniform distribution $\mathcal{U}(0, 1)$\\
            Sample $q_\text{A+R}$ from uniform distribution $\mathcal{U}(0, 1)$\\
            \If{$q_\text{T} < p_\text{T}$}
            {
                Randomly select a tempo ratio $\beta$ from $\{0.9, 1.1\}$ \\
                $\text{segment} = \text{sox\_tempo}(\text{segment}, \beta)$ \\
            }
            \If{$q_\text{A+R} < p_\text{A+R}$}
            {
                Randomly select $\text{noise\_type}$ from $\{\text{additive, reverb}\}$ \\
                $\text{segment} = \text{add\_noise}(\text{segment}, \text{noise\_type})$
            }
        }
    }
\caption{Pseudo code to construct DINO input}
\label{algo:algorithm}
\end{algorithm}
\vspace{-5pt}

The detailed data augmentation and segment sampling pipeline in DINO is shown in Algorithm \ref{algo:algorithm}. In our experiment, we applied the audio perturbation in two different sets up using sox \footnote{\url{http://sox.sourceforge.net/}} toolkit. The first one is ``sox pitch'', which will change the pitch of audio by $\alpha$ `cent' (i.e. 100ths of a semitone). The second one is ``sox tempo'', which will change the tempo (speed) of the audio but keep the pitch. When the audio pitch is changed, we consider the audio from a new speaker. Thus, in Algorithm \ref{algo:algorithm}, pitch perturbation is applied on the whole utterance. After the pitch perturbation augmentation, several long and short segments will be sampled from the utterance. Then, additive noise, reverberation, and tempo perturbation augmentation are applied to each segment independently.

% \begin{table*}[ht!]
% \footnotesize
% \centering
% \caption{\textbf{Results on different evaluation sets.} The supervised training systems are trained with additive noise, reverberation, and speed perturbation (``sox speed'') augmentation.}
% \begin{adjustbox}{width=.8\textwidth,center}
% \begin{threeparttable}
% \begin{tabular}{c|cccccc}
% \hline 
% Model & Vox1-O  & Vox1-E & Vox1-H  & Cnceleb-E  & SITW-dev & SITW-eval \\
% ECAPA-TDNN-S & 3.250 & 3.386 & 6.039 & 11.912 & 6.233 & 5.471 \\
% ECAPA-TDNN-B & 2.505 & 2.473 & 4.791 & 11.467 & 4.757 & 4.544 \\
% Resnet34 & 4.223 & 4.406 & 7.887 & 11.799 & 8.697 & 8.151 \\
% \hline
% $^\dagger$ECAPA-TDNN-S & 1.175 & 1.283 & 2.372 & 10.673 & 2.269 & 2.464 \\
% $^\dagger$ECAPA-TDNN-B & 0.952 & 1.121 & 2.231 & 12.273 & 1.900 & 1.771 \\
% $^\dagger$Resnet34 & 1.048 & 1.141 & 2.113 & 10.194 & 2.023 & 2.041 \\

% \hline
% \end{tabular}
% \begin{tablenotes}\footnotesize
% \item $^\dagger$: supervised training model.            
% \end{tablenotes}
% \end{threeparttable}
% \label{table:res_on_three_eval}
% \end{adjustbox}
% \end{table*}

% \vspace{-10pt}
\section{Experiment Setup}
\label{sec:setup}

\subsection{Dataset}
\label{ssec:dataset_setup}
In our experiment, the Voxceleb2 dev set \cite{chung2018voxceleb2} is used as the training set. When we are doing self-supervised training, we do not use any speaker labels. We sample the audios from MUSAN dataset \cite{musan2015} as the additive noise and use the impulse responses in RIR\footnote{\url{https://www.openslr.org/28/}} as the reverberation augmentation. We evaluate our system on three different evaluation trials 
Vox1-O, Vox1-E, and Vox1-H constructed from Voxceleb1 dataset \cite{nagrani2017voxceleb}.
% We evaluate our system on three different evaluation sets: Voxceleb1 \cite{nagrani2017voxceleb}, Cnceleb \cite{li2022cn,fan2020cn} and SITW \cite{mclaren2016speakers}. There are three public trials for Voxceleb1: Vox1-O, Vox1-E, and Vox1-H. The is one public evaluation set for Cnceleb. For SITW dataset, we evaluated our system on SITW-dev and SITW-eval trials.

\subsection{System Configuration}
\label{ssec:system_configuration}
% In our experiment, we tried both TDNN based ECAPA-TDNN \cite{desplanques2020ecapa} and resnet based r-vector \cite{zeinali2019but} as the speaker embedding extractor. 
In our experiment, we use ECAPA-TDNN \cite{desplanques2020ecapa} as the speaker embedding extractor.
Besides, we denote the  ECAPA-TDNN with the channel number equal to 512 as the ECAPA-TDNN-S (small) and the  ECAPA-TDNN with the channel number equal to 1024 as the ECAPA-TDNN-B (big). Without the specific annotation, the ECAPA-TDNN-S will be used to analyze the characteristics of DINO training strategy.

The projection head in Figure \ref{fig:dino} contains three fully connected layers with hidden dimension 2048 followed by $\ell^{2}$ normalization and a weight normalization fully connected layer \cite{salimans2016weight} which maps the final output to K dimension. In our experiment, we set the K to 65536, the same as the best configuration in \cite{caron2021emerging}. The student temperature $\tau_s$ in equation \ref{eq:get_prob} is set to 0.1. The teacher temperature $\tau_t$ is linearly increased from 0.04 to 0.07 during the first 30 epochs and then fixed. The momentum factor $m$ in equation \ref{eq:c_update} is set to 0.9. Besides, when sampling the segments from one utterance, the long segment is set to 3s and the short segment is set to 2s. The long segments number and short segments number are set by default to 2 and 4, respectively. Cosine scheduler \cite{grill2020bootstrap} is used to optimize the system, the initial learning rate and final learning rate are set to 0.2 and 0.00005, respectively. All the systems are trained for 150 epochs without specific annotation. In each epoch, we iterate all the utterances in the training set.

During the evaluation process, speaker embedding is extracted from the student speaker embedding extractor in Figure \ref{fig:dino}. The cosine score is used as the similarity measurement in our experiment.

\section{Results and Analysis}
\label{sec:result}

\subsection{Augmentation Methods}
\label{ssec:res_data_aug}
\begin{figure}[ht!]
  \centering
  \includegraphics[width=0.47\textwidth]{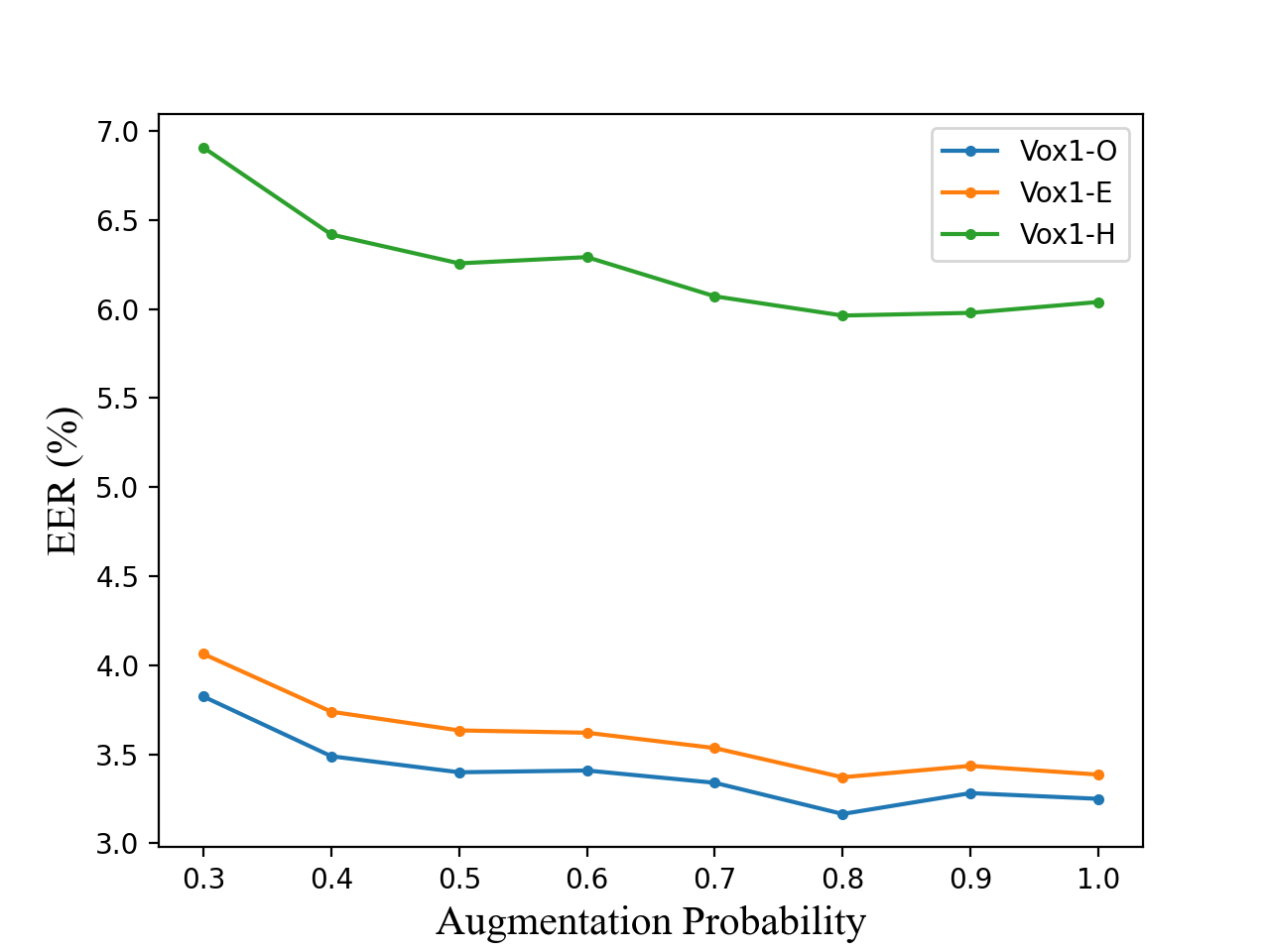}
  \caption{\textbf{Changes of EER (\%) with the probabilities of data augmentation.} Here, only additive noise and reverberation are applied. ECAPA-TDNN-S is used as the embedding extractor.}
  \label{fig:aug_prob}
\end{figure}

In this section, we analyze the effect of different data augmentation methods in DINO training.
We first verify their effectiveness by varying the $p_{A+R}$ in Algorithm \ref{algo:algorithm} and plot the result curve in Figure \ref{fig:aug_prob}. From the figure, we find that extensive augmentation is very important in DINO training. DINO aims to model consistent information within an utterance. This consistent information can be speaker information or channel information. Additive noise and reverberation can break the channel information consistency across different segments from the same utterance and help the model focus on learning speaker information.

\begin{table}[ht!]
\footnotesize
\centering
\caption{\textbf{EER (\%) results with different augmentation methods.} The abbreviations for augmentation methods: A (additive noise), R (reverberation), T (tempo perturbation) and P (pitch perturbation). Different from P shifting the pitch by $200$ or $-200$ cent in Algorithm \ref{algo:algorithm}, for augmentation $\hat{\text{P}}$, the pitch is only shifted by $200$ cent. The probability for each kind of augmentation method are set to: $p_{\text{A}}=1.0$, $p_{\text{R}}=1.0$, $p_{\text{T}}=0.5$, $p_{\text{P}}=0.5$ and $p_{\hat{\text{P}}}=0.5$. The iteration $N$ is the setup introduced in section \ref{ssec:system_configuration}.}
\begin{adjustbox}{width=.48\textwidth,center}
\begin{tabular}{ccc|ccc}

\toprule
Embedding Extractor & Iter \#  & Aug Type & Vox1-O & Vox1-E & Vox1-H \\

\hline
    \multirow{15}{*}{ECAPA-TDNN-S}     &\multirow{9}{*}{N}  & - & 15.22  & 15.28 & 19.95  \\
                    & & A & 4.276 & 4.499 & 7.875 \\ 
                    & & R & 5.212 & 5.626 & 9.100  \\ 
                    & & T & 14.79 & 15.00 & 19.69 \\
                    & & P & 18.49 & 18.30 & 23.53 \\ 
                    & & A + R & 3.250 & 3.386 & 6.039 \\ 
                    & & A + R + T & \textbf{3.212} & \textbf{3.355} & \textbf{5.969} \\ 
             & & A + R + P & 3.856 & 4.003 & 7.535 \\ 
         & & A + R + $\hat{\text{P}}$ & 3.712 & 3.722 & 6.816 \\ 
\cline{2-6}
& \multirow{6}{*}{2N}  & A + R  & 3.271 & 3.172 & 5.773 \\
& & A + R + T & 3.165 & 3.190 & 5.778 \\
& & A + R + P & 3.127 & 3.293 & 6.137 \\
& & A + R + $\hat{\text{P}}$ & 2.989 & \textbf{3.052} & \textbf{5.642} \\
& & A + R + T + P & 3.085 & 3.248 & 6.149 \\
& & A + R + T + $\hat{\text{P}}$ & \textbf{2.968} & \textbf{3.052} & 5.652 \\
\hline
\multirow{8}{*}{ECAPA-TDNN-B}    &\multirow{4}{*}{N}  & A + R & \textbf{2.957} & 3.141 & 5.693 \\
                    & & A + R + T & 2.962 & \textbf{3.044} & \textbf{5.580} \\ 
             & & A + R + P & 3.202 & 3.468 & 6.637 \\ 
          & & A + R + $\hat{\text{P}}$ & 3.064 & 3.124 & 5.768 \\ 
\cline{2-6}
& \multirow{4}{*}{2N}  & A + R  & 2.755 & 2.861 & 5.19 \\
& & A + R + T & 2.718 & 2.781 & 5.141 \\
& & A + R + P & 2.819 & 2.787 & 5.433 \\
& & A + R + $\hat{\text{P}}$ & \textbf{2.505} & \textbf{2.473} & \textbf{4.791} \\
% & & A + R + T + P & - & - & - \\
% & & A + R + T + $\hat{\text{P}}$ & - & - & - \\
\bottomrule
\end{tabular}
\label{table:res_diff_aug}
\end{adjustbox}
\end{table}

To further explore the effect of each data augmentation method, we do an ablation study by adding only one kind of augmentation at a time. Besides, we also explore the audio perturbation augmentation which has been introduced in section \ref{ssec:data_augmentation_intro}.
% is started to be used in speaker recognition task recently \cite{}. We tried two different ways to do speed perturbation using the open source sox \footnote{\url{http://sox.sourceforge.net/}} toolkit. The first one is  ``sox speed'', which can change the utterance speed and pitch. The second one is ``sox tempo'', which can only change the utterance speed. If the pitch is changed, the utterance can be considered from a new speaker. 
We show the results of the ablation study of each augmentation method in Table \ref{table:res_diff_aug}. The observations from the first attempt by using $N$ steps of iterations in the training process are 1) the ECAPA-TDNN-S model trained without data augmentation performs much worse than those with the data augmented by additive noise or reverberation; 2) audio tempo perturbation can improve EER but marginally; 3) audio pitch perturbation degrades the performance. The above observations, i.e. 2) and 3), are the same even when we combine the audio perturbation with the additive noise and reverberation. Such a phenomenon is inconsistent with the results in supervised training \cite{chen2022sjtu}.   

%It is worth noting that the audio pitch perturbation degrades the performance and the audio tempo perturbation only improves the system performance a little. When only applying the audio perturbation augmentation, it is reasonable that there is no obvious improvement. This is because the audio perturbation augmentation does not break the channel information consistency across different segments from the same utterance. However, when we combine the audio perturbation with the additive noise and reverberation, there is still no obvious improvement and the system even degrades when applying audio pitch augmentation. Such a phenomenon is inconsistent with the results in supervised training \cite{chen2022sjtu}. 

% However, the conclusion in section \ref{ssec:res_spk_utt_num} shows that more speakers will make the training more difficult and the ``sox speed'' perturbation will change the pitch of audio, i.e. generating new speakers. The result after `sox speed'' augmentation also further validates previous conclusion in section \ref{ssec:res_spk_utt_num}.

According to our analysis (the details will be shown in section \ref{ssec:res_spk_utt_num}), more speakers in the training set will make the training difficult and thus degrade the performance. The pitch perturbation will change the characteristics of speaker, which is equivalent to generating data with new speakers. To reduce this side effect, we first change the setup of pitch perturbation from $\text{P}$ to $\hat{\text{P}}$, i.e., generating data with fewer new speakers. As expected, the side effect of pitch perturbation augmentation is mitigated a bit indicated by the changes of EER in the Table \ref{table:res_diff_aug}. Considering from another angle, the audio perturbation augmentation can generate more training samples and more training steps are required to make the model converge. So we double the training iterations. The EERs with 2$N$ iterations in the Table \ref{table:res_diff_aug} show that the side effect of pitch perturbation augmentation is further mitigated and the augmentation setup of ``A+R+$\hat{\text{P}}$'' outperforms that of ``A+R''. 

The EER improvements from pitch perturbation are much more significant on the big model, ECAPA-TDNN-B, than those on the small model, ECAPA-TDNN-S. We didn't observe a similar phenomenon in the supervised scenario \footnote{In our supervised training experiment, after pitch perturbation augmentation, the EER (\%) improvement on Vox-H is  0.125 (2.31 to 2.185) for ECAPA-TDNN-B and 0.16 (2.391 to 2.231) for ECAPA-TDNN-S.}, where the small and large model always has similar benefits from the pitch perturbation augmentation. We conjecture that self-supervised approach and large model size can be complementary to each other for speaker representation learning. Self-supervised learning is a more difficult task due to the weaker supervision. Larger models can use the redundant parameters to store more information from training data. Self-supervised learning enables the model size growth to improve the model performance. 

\subsection{Comparing with Other Self-Supervised Learning Methods}
\label{ssec:res_compare}
In this section, we compared our methods with other self-supervised speaker representation learning methods. From the upper part of Table \ref{table:res_compare_with_others}, we notice that the DINO strategy performs the best among all the self-supervised training strategies in speaker verification task. Enhanced with our proposed audio perturbation augmentation strategy, our systems outperform other systems with a very large improvement.

\begin{table}[ht!]
\footnotesize
\centering
\caption{\textbf{Performance comparison of different self-supervised training methods.} All the results are evaluated on the Vox1-O evaluation trial. In this table, our systems are trained with A+R+$\hat{\text{P}}$ augmentation and the iteration number is doubled. For fair comparison with other works, $p_{target}$ in minDCF is set to 0.05 here.}
\begin{adjustbox}{width=.42\textwidth,center}
\begin{threeparttable}
\begin{tabular}{llcc}
\toprule
% \multirow{2}{*}{Methods} & Embedding  & \multirow{2}{*}{EER (\%)} & \multirow{2}{*}{minDCF} \\
% & Extractor & & \\
Methods & Embedding Extractor & EER (\%) & minDCF \\
\hline
AP+AAT \cite{huh2020augmentation} & Fast ResNet34  & 8.65 & 0.4540 \\
MOCO+ProtoNCE \cite{xia2021self} & TDNN  & 8.23 & 0.5900 \\
CEL \cite{mun2020unsupervised} & Fast ResNet34  & 8.01 & N/A \\
Contrastive \cite{tao2022self} & ECAPA-TDNN-S   & 7.36 & N/A \\
SSReg \cite{sang2022self} & Fast ResNet34   & 6.99 & 0.4340 \\
Raw wave DINO \cite{jung2022pushing} &  RawNet3  & 5.40 & 0.3396 \\
DINO + CL \cite{heo2022self} & ECAPA-TDNN  & 4.47 & 0.3057 \\
\hline
% \multirow{3}{*}{Ours} & ResNet34 & 3.665 & 0.3683 \\
\multirow{2}{*}{Ours} & ECAPA-TDNN-S & 2.989 & 0.2113 \\
 & ECAPA-TDNN-B & \textbf{2.505} & \textbf{0.1626} \\

% AP+AAT \cite{huh2020augmentation} & Fast ResNet34  (1.4M)  & 8.65 & 0.4540 \\
% MOCO+ProtoNCE \cite{xia2021self} & TDNN  (4.5M)  & 8.23 & 0.5900 \\
% CEL \cite{mun2020unsupervised} & Fast ResNet34  (1.4M)  & 8.01 & N/A \\
% Contrastive \cite{tao2022self} & ECAPA-TDNN-S  (6.2M)  & 7.36 & N/A \\
% SSReg \cite{sang2022self} & Fast ResNet34  (1.4M)  & 6.99 & 0.4340 \\
% Raw wave DINO \cite{jung2022pushing} &  RawNet3 (16.3M)  & 5.40 & 0.3396 \\
% DINO + CL \cite{heo2022self} & ECAPA-TDNN  & 4.47 & 0.3057 \\
\bottomrule
\end{tabular}
% \begin{tablenotes}\footnotesize
% \item $^*$: $p_{target}$ for minDCF is set to 0.05
% \end{tablenotes}
\end{threeparttable}
\label{table:res_compare_with_others}
\end{adjustbox}
\end{table}

\subsection{Fine-tuning on Small Amount of Labeled Data}
\label{ssec:res_dino_finetune}
In many scenarios, the training process by leveraging a massive amount of unlabeled data is always considered as a pre-training stage and is usually followed by a fine-tuning stage where a small amount of labeled data from a specific domain is used for continuous training. In this section, we test how our system performs when we consider it as a pre-training model. We fine-tune our systems on the Voxceleb1 dev set and show the result on trial Vox1-O in Table \ref{table:res_finetune}. From the table, we find that our best system outperforms most of the other systems. Besides, our best system has a comparable performance with Unispeech-SAT Large model \cite{chen2022large}, which has been trained on 
tens of thousands of hours audio data and has a very large model size.

\begin{table}[ht!]
\footnotesize
\centering
\caption{\textbf{EER (\%) results after fine-tuned on Voxceleb1 dev set.} The model is fine-tuned on Voxceleb1 dev set supervised by AAM loss \cite{deng2019arcface}. Data augmentation with additive noise and reverberation is applied in the fine-tuning. All the results are evaluated on Vox1-O evaluation trial.}
\begin{adjustbox}{width=.48\textwidth,center}
\begin{threeparttable}
\begin{tabular}{cccc}
\toprule 

 & \multirow{2}{*}{Methods (Model Param \#)} & Amount of & \multirow{2}{*}{Vox-O} \\
 & & Pre-training Data & \\
\hline
\multirow{2}{*}{Chen et al. \cite{chen2022large}} & UniSpeech-SAT Base ($\sim 100$M) & 94k hrs & $^*$1.611 \\
           & UniSpeech-SAT Large ($\sim 320$M) & 94k hrs & $^*$1.218 \\
Zhang et al. \cite{zhang2021contrastive}  & AP + Channel-Invariant  (1.4M)       &  2.36k hrs & 3.88  \\
Jung et al. \cite{jung2022pushing}  & Raw wave DINO (16.3M)           &  2.36k hrs & 2.18  \\
Heo et al. \cite{heo2022self}       & DINO + CL           &  2.36k hrs& 1.84 \\
\hline
\multirow{4}{*}{Ours}& ECAPA-TDNN-S (6.2M) & No Pre-training &  2.484 ($^*$2.234) \\
                    % & $^\dagger$ ECAPA-TDNN-S (6.2M) & No Pre-training &  2.026 ($^*$1.840) \\
                    & ECAPA-TDNN-B (14.7M) & No Pre-training & 2.356 ($^*$2.180)\\
                    % & $^\dagger$ECAPA-TDNN-B (14.7M) & No Pre-training & 1.787 ($^*$1.622)\\
& ECAPA-TDNN-S (6.2M) & 2.36k hrs &  1.702 ($^*$1.505) \\
                    % & $^\dagger$ ECAPA-TDNN-S (6.2M) & 2.36k hrs &  1.558 ($^*$1.393) \\
                    & ECAPA-TDNN-B (14.7M) & 2.36k hrs & 1.335 ($^*$1.228)\\
                    % & $^\dagger$ECAPA-TDNN-B (14.7M) & 2.36k hrs & 1.191 ($^*$1.154)\\
\bottomrule
% & \multirow{2}{*}{Frame Work} & \multirow{2}{*}{Model Param \#} & Supervised Training Data & \multirow{2}{*}{Vox-O EER (\%)} \\
%  & & & (no vox1 finetuning) & \\
%  \hline
% \multirow{2}{*}{Desplanques et al. \cite{desplanques2020ecapa}} & ECAPA-TDNN-S & 6.2M & 2.36k hrs (Vox2 dev) &  $^*$1.01 \\
%  & ECAPA-TDNN-B & 14.7M & 2.36k hrs (Vox2 dev) &  $^*$0.87 \\
% \hline

\end{tabular}
\begin{tablenotes}\footnotesize
\item $^*$: doing the adaptive score normalization in the scoring process
% \item $^\dagger$: applying extra pitch perturbation $P$ in the fine-tuning process
\end{tablenotes}
\end{threeparttable}
\label{table:res_finetune}
\end{adjustbox}
\end{table}

\subsection{The number of Speakers and Utterances in the Training Corpus}
\label{ssec:res_spk_utt_num}
In this section, we explore the impact of the number of speakers and utterances in the training set on the DINO self-supervised training strategy. Here, we randomly sample a small set from the original Voxceleb2 dev set in two different ways. In the first way, we sample the utterances according to the speaker identity. In the second way, we just randomly sample a small set from the original training set.
Although we have no way to know how many people are in the unlabeled audio data, such an analysis can help us better understand the self-supervised training algorithm and guide how we collect data.

We list the corresponding results in Table \ref{table:res_spk_num_utt_num}. In the table, we take a half or a quarter subset from the original training data. As said above, we sample the subset according to speaker identity or directly randomly sampled each utterance. Interestingly, for two subsets with the same amount of utterances, the subset with more speakers performs worse. This phenomenon seems counter-intuitive and supervised training system shows the opposite trend. We speculate that DINO needs to traverse enough audios belonging to a particular person to model that person. For two subsets with the same number of voices, the model may not be able to model each speaker well when there are fewer utterances belonging to each speaker. In contrast, supervised training may focus on discriminating more speakers. Besides, Heo et al. \cite{heo2022self} also found a similar phenomenon that reducing the speaker number in the DINO training can make the task easier and they proposed a curriculum learning strategy to improve the self-supervised learning strategy. 

\begin{table}[ht!]
\footnotesize
\centering
\caption{\textbf{EER (\%) results of systems trained on the dataset with different speakers and utterances.} $N_{spk}$ and $N_{utt}$ denote the total speaker number and total utterance number in Voxceleb2 development set. The ECAPA-TDNN-S is used as the embedding extractor. Additive noise and reverberation augmentation are applied in the training process with $p_{A+R}=1.0$. We put the results of the supervised system in the bracket. The supervised system is trained with AAM loss \cite{deng2019arcface} and we apply the additive noise and reverberation augmentation in the training process.}
\begin{adjustbox}{width=.42\textwidth,center}
\begin{threeparttable}
\begin{tabular}{cc|ccc}
\toprule
Speaker \# & Utt \# & Vox1-O  & Vox1-E  & Vox1-H  \\
\hline
$N_{spk}$                   & $N_{utt}$                 & 3.250 (1.117) & 3.386 & 6.039 \\

$\slfrac{N_{spk}}{2}$       & $\slfrac{N_{utt}}{2}$     & 3.516 (1.744) & 3.598 & 6.425 \\
$N_{spk}$                   & $\slfrac{N_{utt}}{2}$     & 4.526 (1.510) & 4.697 & 8.022 \\
$\slfrac{N_{spk}}{4}$       & $\slfrac{N_{utt}}{4}$     & 4.643 (2.803) & 4.696 & 7.968 \\
$N_{spk}$                   & $\slfrac{N_{utt}}{4}$     & 4.936 (2.026) & 5.338 & 8.979 \\
\bottomrule
\end{tabular}

\end{threeparttable}
\label{table:res_spk_num_utt_num}
\end{adjustbox}
\end{table}

\subsection{Probing the Characteristics of DINO Output Distribution}
\label{ssec:res_dino_output_characteristic}
As introduced in section \ref{ssec:dino_system_intro}, the DINO is trained in a teacher-student paradigm. The model is optimized to match the output distribution from student and teacher networks. In our experiment, we find that there is one dimension dominating the teacher output distribution, which is very similar to a one-hot vector used in the classification-based cross-entropy loss. We're wondering if it is possible to consider the dimension index of the largest value in the distribution as some kind of pseudo speaker label. In Table \ref{table:res_nmi}, we calculate the normalized mutual information (NMI) \cite{guimera2007module} between true speaker labels and estimated pseudo labels on the Voxceleb2 dev set. We consider the index of the largest value or the indexes combination of the top two values in the teacher output distribution as the pseudo label. From the results, we find the pseudo label has a strong correlation with the true label. Besides, the index of the second largest value couldn't further improve the NMI.

\begin{table}[ht!]
\footnotesize
\centering
\caption{\textbf{Normalized mutual information (NMI) between estimated speaker label from DINO output distribution and true speaker labels.} 
The NMI value ranges from 0 to 1 and larger value represent stronger correlation. Speaker \# denotes the pseudo speaker label number. In this table, the systems are trained with A+R+$\hat{\text{P}}$ augmentation and the iteration number is doubled. 
% Argmax Top 1 means we use the index of the largest value of DINO distribution as the pseudo label and the maximum number of the pseudo class is K. Argmax Top 2 means we use the indexes combination of the largest two values of DINO distribution as the pseudo label and the maximum number of the pseudo class is $\text{K}(\text{K}-1)$.
}
\begin{adjustbox}{width=.47\textwidth,center}
\begin{tabular}{c|cc|cc}
\toprule
\multirow{2}{*}{Backbone}  & \multicolumn{2}{c}{Argmax Top 1} & \multicolumn{2}{c}{Argmax Top 2}\\
\cline{2-5}
                    & NMI & Pseudo Speaker \# & NMI & Pseudo Speaker \# \\
\hline
ECAPA-TDNN-S & 0.816 & 2595 & 0.806 & 24725 \\
ECAPA-TDNN-B & 0.790 & 1319 & 0.785 & 17900 \\
% Resnet34 & - & - & - & - \\

\bottomrule
\end{tabular}
\label{table:res_nmi}
\end{adjustbox}
\end{table}

\vspace{-10pt}
\subsection{The Number of Segments}
\label{ssec:res_segment_strategy}
In section \ref{ssec:dino_system_intro}, we have introduced the multi-segment strategy, where multiple long and short segments are sampled from the same utterance.
Here, we explore the importance of this strategy.
The corresponding results are listed in the Table \ref{table:res_segment_number}.
From the results, we find that increasing the number of long and short segments can consistently improve performance.
Because the average utterance duration of Voxceleb2 is 8s and more segments will cost more computation, we only sample at most two long segments and four short segments.

\begin{table}[ht!]

\footnotesize
\centering
\caption{\textbf{EER (\%) results for different segmentation number setups.} The ECAPA-TDNN-S is used as the embedding extractor. Additive noise and reverberation augmentation are applied in the training process with $p_{A+R}=1.0$.}
\begin{adjustbox}{width=.4\textwidth,center}
\begin{tabular}{cc|cccc}
\toprule
Long Seg \# & Short Seg \# & Vox1-O & Vox1-E & Vox1-H\\
\hline
1 & 1 & 4.074 & 4.588 & 8.665 \\
1 & 2 & 3.516 & 3.730 & 6.827 \\
2 & 4 & \textbf{3.250} & \textbf{3.386} & \textbf{6.039} \\
\bottomrule
\end{tabular}
\label{table:res_segment_number}
\end{adjustbox}

\end{table}

\vspace{-10pt}
\section{Conclusion}
In this paper, we presented a DINO based self-supervised framework for speaker representation learning. The speaker pseudo labels predicted from the output distributions of our self-supervised model have a strong correlation with the true labels. Our ablation studies also indicate that data augmentation with the proposed audio perturbation strategy can further improve the model performance in addition to additive noise and reverberation. In the future, we plan to increase the data size by leveraging the unlabeled data in the wild, in which we haven't achieved a better performance by using the current settings. We believe we also need to scale up the model size accordingly. 
\bibliographystyle{IEEEbib}
\bibliography{strings,refs}

\end{document}